\documentclass[11pt]{article}
\usepackage{epsfig}
\usepackage[usenames]{color}
\usepackage{amsmath}
\usepackage{amsfonts}
\usepackage{amssymb}
\newtheorem{proposition}{Proposition}[section]
\newcommand{\bpr}{\begin{proposition}}
\newcommand{\epr}{\end{proposition}}

\newcounter{Roman}
\addtocounter{Roman}{1}

\newcommand{\beq}{\begin{equation}}
\newcommand{\eeq}{\end{equation}}
\newcommand{\bea}{\begin{eqnarray}}
\newcommand{\eea}{\end{eqnarray}}
\newcounter{saveeqn}

\newcommand{\D}{\displaystyle}

\newcommand{\hj}{\hat{\jmath}}

\newcommand{\bx}{{\bf x}}
\newcommand{\tr}{{\rm tr}}

\newcommand{\re}{{\rm Re}}

\newcommand{\vev}[1]{\Big\langle #1 \Big\rangle}
\newcommand{\bpsi}{\bar{\psi}}

\newcommand{\C}{{\cal C}}

\input epsf
\begin{document}  

\begin{center}{\Large\bf  Chiral symmetry in  
$SU(N_c)$ gauge theories at high density }\\[2cm] 
{E. T. Tomboulis\footnote{\sf e-mail: tomboulis@physics.ucla.edu}
}\\
{\em Department of Physics and Astronomy, UCLA, Los Angeles, 
CA 90095-1547}
\end{center}
\vspace{1cm}

\begin{center}{\Large\bf Abstract}\end{center} 
We study $SU(N_c)$ lattice gauge theories with $N_f$ flavors of massless staggered  
fermions  in the presence of 
quark chemical potential $\mu$. A recent exact result that in the strong coupling limit (vanishing inverse gauge coupling $\beta$) and for sufficiently large $\mu$ the theory is in a chiral symmetric phase is here extended into the finite gauge coupling region. A cluster expansion combining a fermion spacelike hopping expansion and a strong coupling plaquette expansion is shown to converge for sufficiently large $\mu$ and small $\beta$ at any temperature $T$. All expectations of chirally non-invariant local fermion operators vanish identically, or, equivalently, their correlations cluster exponentially within the expansion implying absence of spontaneous chiral symmetry breaking. The resulting phase at low $T$ 
may be described as a ``quarkyonic" matter phase.   Some implications for the phase diagram of $SU(N_c)$ theories are discussed. 

\vspace{5cm}

\begin{center}
{  \Large  In memory of Pierre van Baal }
\end{center}

\vfill
\pagebreak

\section{Introduction} 
\setcounter{equation}{0}
\setcounter{Roman}{0}

The phase diagram of QCD, and more generally that of $SU(N_c)$ gauge theories, as a function of temperature, density and number of fermion flavors remains a major challenge. In the presence of a finite chemical potential the fermion determinant becomes complex (the ``sign problem") and this presents application of standard Monte-Carlo techniques. This stumbling block has hindered general exploration of the phase diagram, in particular at intermediate and large quark chemical potential. 
Despite progress in recent years, made by a combination of numerical simulations, mostly for small chemical potential,  analytical techniques and investigation of model systems (see \cite{FH} \cite{dF} for some review), the QCD phase diagram away from the region along the temperature axis 
remains, for the most part, not firmly established. More generally, the phase diagram of $SU(N_c)$ 
theories with varying fermion content, of great current interest for physics beyond the Standard Model, is also largely unexplored. 

In a recent paper \cite{T1} we showed that $SU(N_c)$ lattice gauge theory with $N_f$ flavors of massless staggered  fermions  in the strong coupling limit, i.e. at vanishing inverse gauge bare coupling $\beta$,  is in a chirally symmetric phase provided the quark chemical potential $\mu$ is large enough. This is an exact result obtained by means of a convergent cluster expansion. Here we extend this result into the region of non-vanishing $\beta$.

The theory in the strong coupling limit in the presence of chemical potential has been investigated in the literature, mostly for the cases of $SU(2)$ and $SU(3)$ and  $N_f=1$, in a variety of approaches. One approach relies on 
integrating out the gauge field in the strong coupling limit. This results in a representation of the partition function in terms of monomers, dimers and baryon loops \cite{DMW}, or monomers, dimers and polymers 
\cite{KM}, \cite{KIM}. The sign problem is partly evaded within this representation, thus allowing simulations. In such simulations in the case of $N_c=3$ \cite{KM} a chiral symmetry restoring first order  transition was found at some critical $\mu$. Similarly, for $N_c=2$, restoration of chiral symmetry at large $\mu$ and/or $T$ was seen in  \cite{KIM}.  
More recently, the two-color ($N_c=2$), $N_f=1$ case was investigated in \cite{CJ} using the dimer-baryon loop representation with a new updating algorithm \cite{AC}. A second order transition to a chirally symmetric phase at some critical $\mu$ was seen in good agreement with  mean field predictions.  In the case of $N_c=3$ such improved simulations were carried out in \cite{FdF}. Another approach is based on mean field investigations of effective actions.  In  \cite{NFH}, \cite{N}, \cite{KMOO} an effective action was obtained by performing a $1/d$ expansion in the spatial directions and retaining only the leading terms, while leaving the timelike directions intact. 
This effective action was used to obtain the phase diagram in $T$, $\mu$ and quark mass $m$ within  
a mean field approximation. This phase diagram exhibits a chiral phase at large $\mu$ and/or $T$.  The existence of this phase for general $SU(N_c)$ was finally rigorously proven in \cite{T1} by means of a cluster expansion shown to converge for large $\mu$ in the infinite volume limit.

In this paper we extend this result into the region of non-zero $\beta$, i.e. we show that the theory is in a chirally symmetric phase for sufficiently large quark chemical potential $\mu$ and small $\beta$
at any $T$. 
This is done by means of a cluster expansion which combines the fermion hopping expansion on spacelike bonds employed in \cite{T1} with a polymer-type expansion of the gauge field plaquette interaction. 
The expansion is shown to converge in the infinite volume limit. The existence of chiral symmetry 
follows then as a straightforward consequence of this convergence. The result holds for any choice of $N_c, N_f$. 
The resulting physical picture is that of a strongly interacting chirally symmetric gauge theory, a system with the qualitative features at low $T$ of what has been dubbed  ``quarkyonic"  matter \cite{McLP}. 
Implications for the phase diagram of the theory, in particular in connection with the existence 
of color superconductivity at large $\mu$, are discussed in the section 4 below.

The paper is organized as follows. In section 2 the formalism of lattice gauge theories at finite 
quark chemical potential is cast in a form appropriate for the cluster expansion presented 
in section 3.  (It was found  necessary to include some replication of the formalism in \cite{T1} in order to make the paper self-contained.) The expansion is set up and its diagrammatic structure laid out in section 3.1. We then formulate it as a polymer expansion of the logarithm of the partition function and of expectations of observables. Its convergence is examined in section 3.2. The preservation of chiral symmetry within the expansion follows then as a standard consequence of its convergence (section 3.3). Discussion of the result and outlook are given in section 4.

\section{The $SU(N)$ lattice gauge theory at finite chemical potential  \label{S1}}
\setcounter{equation}{0}
\setcounter{Roman}{0}

We work on a $(d+1)$-dimensional periodic hypercubic lattice $\Lambda$ of size $L_s^d \times L$
and lattice spacing $a$. The lattice lengths in Euclidean time ($L$) and space ($L_s$) are taken to be even. 
The physical temperature is then $T=(La)^{-1}$. 
The spatial lattice obtained as a particular equal time slice of $\Lambda$ will be denoted by $\Lambda_s$. 
Lattice site coordinates will be 
denoted by $x=(x^\lambda)= (x^0, {\bf x})$ with $\lambda=0,1,\ldots, d$, and 
${\bf x}=(x^j)$, $j=1,\ldots,d$. The alternate notation  
$x^0=\tau$ is also used. Lattice unit vectors in the space and time directions will be denoted by 
$\hat{\jmath}$ and $\hat{0}$, respectively. We generically denote lattice bonds by $b$, plaquettes by $p$, etc.  Bonds will be specified  more explicitly 
as $b=(x, \lambda)$, or $b_s=(x, j)=<x, x+\hat{\jmath}>$ if spacelike, and $b_\tau=(x, 0)=<x, x+\hat{0}> $ if timelike. Correspondingly, the gauge field variables $U_b$ defined on each $b\in \Lambda$ will, as usual, often be more explicitly specified by $U_j(x)$ and $U_0(x)$. 

The gauge fields $U_b\in G_c$ are taken to transform in the fundamental representation
of the gauge color group $G_c=SU(N_c)$. The fermions $\bpsi_\alpha(x)$, $\psi_\alpha(x)$, ($\alpha=1,\ldots,\nu$), the $2\nu$ generators of a Grassmann algebra on each site, transform as $N_f$ copies (flavors) of the fundamental representation of the gauge group.  
We use $a=1,\ldots, N_c$, $i=1,\ldots,N_f$ for color and flavors indices, respectively. We assume periodic boundary conditions for the gauge field and antiperiodic boundary conditions for the fermions.

The lattice action is 
\beq
S= S_g + S_F  \label{act0}
\eeq 
where 
\beq 
S_g= \sum_p \beta\,[{1\over N_c}\re \,\tr U_p   - 1] \; ,  \label{act1} 
\eeq
with $U_p=\prod_{b\in\partial p} U_b$, is the Wilson gauge field action at inverse gauge coupling $\beta$. 
The fermion action for massless fermions in the presence of chemical potential $\mu/ a$ 
is given by: 
\beq 
S_F = \sum_{x,y} \bpsi(x)  {\bf M}^{(s)}_{xy}(U) \psi(y)  + 
 \sum_{x,y} \bpsi(x) {\bf M}^{(t)}_{xy}(U)  \psi(y) \, ,\label{act2}  
\eeq 
where the matrices ${\bf M}^{(s)}$ and ${\bf M}^{(t)}$ have nonvanishing elements 
only between nearest-neighbor sites, i.e., on bonds, given by
\beq
{\bf M}^{(s)}_{x (x+\hat{\jmath})}(U) = {1\over 2} (a_\tau/ a_s ) 
  \gamma_j(x) U_j(x)  \qquad \mbox{and} \qquad   
  {\bf M}^{(s)}_{(x+\hat{\jmath})x}(U) = -{1\over 2} 
  (a_\tau/ a_s)  \gamma_j (x)U^\dagger_j(x)  \label{Ms}
  \eeq 
for spacelike neighbors, and 
\beq 
{\bf M}^{(t)}_{x, x+\hat{0}}(U) = {1\over 2} \gamma_0 e^{\mu } U_0(x)
\qquad \mbox{and} \qquad {\bf M}^{(t)}_{x+\hat{0},x}(U) = -{1\over 2} \gamma_0 e^{-\mu } 
U^\dagger_0(x) \label{Mt} 
\eeq
for timelike neighbors. 
The matrices $\gamma[b]=\gamma_\lambda(x)$ defined on each bond $b=(x,\lambda)$ 
satisfy $\prod_{b\in \partial p}\gamma[b]=1$ for each plaquette $p$. 
For staggered fermions 
\beq
\gamma_\lambda(x) = (-1)^{\sum_{\nu< \lambda} x^\nu}  \, ,   \qquad \gamma_0(x) = 1 \, .\label{gamma1} 
\eeq
For staggered fermions then one has $\nu=N_cN_f$, and thus $\alpha=(a,i)$. Recall that $N_f$ staggered flavors correspond to $4N_f$ continuum flavors. 
We only consider staggered fermions in this paper even though we often state formulas for general $\gamma[b]$'s. 

With $N_f$ flavors of staggered fermions the fermion action in (\ref{act2}) possesses a chiral 
$U(N_f)\times U(N_f)$ global symmetry of independent rotations of fermions on even and odd sublattices. Explicitly, for any element $(u,v)\in U(N_f)\times U(N_f)$, 
they are given by $\psi(x) \to u\psi(x)$ and $\bpsi(x) \to \bpsi(x)v^\dagger$ for even sites 
and $\psi(x) \to v \psi(x)$ and $\bpsi(x) \to \bpsi(x)u^\dagger$ for odd sites. 

 The full measure is then  
\beq 
d\mu_\Lambda = Z_\Lambda^{-1} \,\prod_{b\in \Lambda} dU_b \prod_{x\in \Lambda} d\bpsi(x) d\psi(x) \exp(S_g + S_F) \, , \label{meas1}
\eeq
with $dU_b$ denoting normalized Haar measure on the group $G_c$, and 
$d\bpsi(x) d\psi(x)\equiv \prod_\alpha d\bpsi_\alpha(x) d\psi_\alpha(x)$ the standard Grassmann algebra measure. The partition function $Z_\Lambda$ is defined by $\int d\mu_\Lambda=1$. 
Expectations of general fermionic observables ${\cal O}$ are given by 
\beq 
\vev{{\cal O}} = \int d\mu_\Lambda \; {\cal O}  \,. \label{exp1}
\eeq

The presence of a nonvanishing chemical potential in (\ref{act2}) introduces an anisotropy between the spacelike and timelike directions which can be exploited to set up a convergent expansion for large $\mu$. (This is analogous to using the corresponding anisotropy for an expansion at large $T$ \cite{TY}.) To do this we will 
rewrite the measure (\ref{meas1}) in a form that will facilitate such an expansion.  
For each spacelike bond $b_s=<x,x+\hj>$ and each $\alpha$ let 
\beq
f_{b_s}^{1,\alpha} \equiv 
\bpsi_\alpha(x) ({\bf M}^{(s)}_{x (x+\jmath)}\psi(x+\jmath) )_\alpha \;, 
\qquad 
f_{b_s}^{2, \alpha} \equiv (\bpsi(x+\jmath) {\bf M}^{(s)}_{(x+\jmath)x})_\alpha \psi_\alpha(x)  
\;. \label{fdef}    
\eeq
Note that $(f_{b_s}^{l,\ \alpha})^2=0$. One then has 
\beq 
\exp \left( 
\bpsi(x) {\bf M}^{(s)}_{x(x+\hj)}(U) \psi(x+\hj) + 
 \bpsi(x+\jmath) {\bf M}^{(s)}_{(x+\hj)x}(U) \psi(x)\right) 
= \  
\prod_{l=1}^2 \prod_{\alpha=1}^{\nu} (1 +  f_{b_s}^{l, \alpha})  \; . \label{frep} 
\eeq

For each site $\bx$ in a fixed time slice $\Lambda_s$ let 
\beq 
d\mu_{\bx}  =  {1\over z } \prod_{\tau=1}^L dU_0(\tau, \bx) 
d\bpsi(\tau,{\bx}) d\psi(\tau,{\bx})
\exp\left( \sum_{\tau, \tau^\prime} \bpsi(\tau,\bx) {\bf M}^{(t)}_{(\tau,\bx) (\tau^\prime,\bx)}(U_0)  \psi(\tau^\prime,\bx)  \right)  \,. 
\label{dmux}
\eeq
Note that 
\bea
& &  
 \sum_{\tau, \tau^\prime} \bpsi(\tau,\bx) {\bf M}^{(t)}_{(\tau,\bx) (\tau^\prime,\bx)}(U_0)  \psi(\tau^\prime,\bx)     \nonumber \\
& = &  
\sum_{\tau=1}^L[ \bpsi(\tau,\bx) \gamma_0 e^{\mu }U_0(\tau,\bx)\psi(\tau+1,\bx) - 
\bpsi(\tau+1,\bx) \gamma_0 e^{-\mu}U^\dagger_0(\tau,\bx)\psi(\tau,\bx) ] \label{dmuxact1}
\eea
with $\psi(L+1,\bx) \equiv - \psi(1,\bx)$, $\bpsi(L+1,\bx) \equiv - \bpsi(1,\bx)$ due to the antiperiodic fermion boundary conditions. 
In (\ref{dmux}) the factor $z$ is defined by $\int d\mu_{\bx}=1$. It represents the  
partition function for a  1-dimensional timelike fermion chain, which, upon integrating out the fermions, is given by 
\beq
z = \int \prod_{\tau=1}^L dU_0(\tau, \bx) \, {\rm Det}{\bf M}^{(t)}_\bx(U_0) 
\label{z}
\eeq
with ${\bf M}^{(t)}_{\bx}(U_0) \equiv \left( {\bf M}^{(t)}_{(\tau,\bx) (\tau^\prime,\bx)}(U_0)\right)$  denoting the restriction of ${\bf M}^{(t)}_{xy}(U_0)$ 
to the submatrix of timelike bonds at fixed $\bx$.

We also expand the exponential of the plaquette gauge field action in characters:
\beq
e^{{\beta\over N_c }\re \,\tr U_p }  =  \sum_j d_j a_j(\beta) \chi_j(U_p)  \, . \label{chexp1}
  \eeq
Here $\chi_j$ denotes the character of dimension $d_j$ in a complete set of irreducible representation characters of the gauge group enumerated by $j$. It is more convenient to factor out the trivial character and work in terms of normalized expansion coefficients:  
\beq  c_j(\beta) =a_i(\beta)/a_0(\beta) \, , \qquad 0\leq c_j \leq 1 \;, 
\eeq 
and write 
\bea 
e^{{\beta\over N_c }\re \,\tr U_p }  & = & a_0(\beta) \, \Big[ 1 + \sum_{j\not= 0} d_j c_j(\beta) \chi_j(U_p) \Big]   \\
   & \equiv  &   a_0(\beta) \, \Big[ 1 + f_p(U_p) \Big] \label{chexp2}
  \eea
E.g., for $G_c=SU(2)$, one has $c_j(\beta) = I_{2j+1}(\beta)/I_1(\beta)$, $j=0,1/2, 1,\ldots$.  
Also, we define the quantity 
\beq 
||f_p|| = \sum_{j\not= 0} d_j^2 c_j(\beta) \label{fnorm}
\eeq
which bounds $f_p(U_p)$ for all $U$ from above (its $||\cdot||_\infty$ norm).

Using (\ref{frep}), (\ref{dmux}) and (\ref{chexp2}), and 
dropping the inessential overall constant factors $e^{-\beta}a_0$,  
the full measure (\ref{meas1}) can now be expressed in the form
\beq 
d\mu_\Lambda = \tilde{Z}_\Lambda^{-1} \prod_{\bx \in \Lambda_s} d\mu_{\bx} 
\prod_{b_s\in \Lambda} dU_{b_s} \prod_{p\in \Lambda} [ 1 + f_p(U_p) ]
 \prod_{b_s\in \Lambda} 
 \prod_{l=1}^2  \prod_{\alpha=1}^{\nu} (1 +  f_{b_s}^{l, \alpha}) \, \label{meas2}
\eeq
with 
\beq 
\tilde{Z}_\Lambda \equiv Z_\Lambda \slash z^{|\Lambda_s|}  \,. \label{tildeZ} 
\eeq
The fermionic timelike part of the measure (\ref{meas2}) factorizes in a product with each factor representing the fermionic degrees of freedom coupled in a 1-dimensional timelike chain at fixed spatial coordinates $\bx$.  It is then very expedient to adopt the so-called Polyakov gauge where the bond variables $U_0(\tau, \bx)$ are independent of $\tau$ and diagonal: 
\bea
U_0(\tau, \bx) & = & {\rm diag} (e^{i\theta_1(\bx)/L}, e^{i\theta_2(\bx)/L}, \cdots, 
e^{i\theta_{N_c}(\bx)/L} ) 
\; . \label{Polgauge1}\\
& = & \exp (i\Theta(\bx)/L)   \label{Polgauge2}
\eea
with 
\beq 
\sum_{a=1}^{N_c} \theta_a(\bx) = 0   \qquad \mbox{and } \qquad  \Theta(\bx) \equiv {\rm diag}( 
\theta_1(\bx), \theta_2(\bx), \cdots, \theta_{N_c}(\bx)) \label{Polgauge3} \,.
\eeq
After a time Fourier transform the 
timelike action  (\ref{dmuxact1})  becomes 
\beq 
 \sum_{\tau, \tau^\prime} \bpsi(\tau,\bx) {\bf M}^{(t)}_{(\tau,\bx) (\tau^\prime,\bx)}(U_0)  \psi(\tau^\prime,\bx)  = 
 \sum_{k_m \in {\sf BZ}} \bpsi(k_m,\bx) i \gamma_0 \sin \Big( k_m + 
 \Theta(\bx)/L -i\mu  \Big) \psi(k_m, \bx) \,. \label{dmuxact2}
 \eeq
The  timelike propagator in the background of the gauge field (\ref{Polgauge2}) then is 
\beq 
C_{\tau, \tau^\prime}( \Theta(\bx)) \equiv C(\tau - \tau^\prime, \Theta(\bx)) = {1\over L} \sum_{k_m\in {\sf BZ}}e^{\D i k_m (\tau - \tau^\prime)}  C(k_m, \Theta(\bx) ) 
\label{prop1}
 \eeq 
 with  
 \beq
 C(k, \Theta(\bx) ) =  \left[ i\gamma_0  \sin \Big( k +  \Theta(\bx)/L -i\mu  \Big)\right]^{-1} \, . 
 \label{prop2}
 \eeq
In (\ref{dmuxact2}) and (\ref{prop1})  the summation is over momenta in the Brillouin zone ({\sf BZ}): 
 \beq 
 k_m = (2m-1) \pi / L \;, \quad  -L/2 + 1 \leq  m  \leq L/2  \qquad \Rightarrow  \quad 
 -\pi + \pi/L \leq k_m \leq \pi - \pi/L    \label{BZ} 
 \eeq 
(integer $m$). Evaluation of (\ref{prop1}) \cite{T1} gives 
\bea
C(\tau - \tau^\prime, \Theta(\bx))_{ai,bj} & = & \delta_{ab}\delta_{ij} [1 - (-1)^{\D (\tau-\tau^\prime)}] \, 
{  e^{\D -i\theta_a(\bx) (\tau-\tau^\prime)/L}  \over 
1 +e^{\D -i\theta_a(\bx) }  e^{ \D - \mu L}  } \, e^{\D - \mu (\tau - \tau^\prime)} \nonumber \\
& &\nonumber \\
& &  \quad \qquad \qquad \qquad\qquad  \mbox{for} \quad
  (\tau-\tau^\prime) > 0, \quad \mu >0 
\label{prop3a}
\eea
and 
\bea
 C(\tau - \tau^\prime, \Theta(\bx))_{ai,bj} 
& = & \! -\delta_{ab}\delta_{ij} [1 - (-1)^{\D |\tau-\tau^\prime|}] \, 
{e^{\D -i\theta_a(\bx)[1 -  |\tau-\tau^\prime|/L]} 
 \over  1 +e^{\D -i\theta_a(\bx) } e^{\D  - \mu L}  } \, 
 e^{ \D - \mu[L -  |\tau - \tau^\prime|]} \nonumber \\
 & & \nonumber \\
 & & \qquad  \qquad \qquad \qquad \qquad \qquad 
\mbox{for} \quad
  (\tau-\tau^\prime) < 0, \quad \mu >0 \; .
\label{prop3b}
\eea
Note that, for $\tau^\prime > \tau$, propagating backward in time from $\tau^\prime$ to $\tau$ is equivalent to propagating forward from $\tau^\prime$ winding around the periodic time direction  to $\tau$.  

For $\mu < 0$ (i.e., nonvanishing antiquark chemical potential)  replace $\mu, \theta_a(\bx)$ by $|\mu|, -\theta_a(\bx)$,  and reverse the sign condition on $(\tau - \tau^\prime)$ in (\ref{prop3a}) - (\ref{prop3b}).  

As seen from (\ref{prop3a}) - (\ref{prop3b}),  $C(\tau, \Theta(\bx))$  vanishes for even $\tau$. This is a consequence of the chiral invariance of the action. The other salient property of $C(\tau, \Theta(\bx))$ is its 
exponential decay for nonvanishing $\mu$. 

The quantity ${\rm Det}{\bf M}^{(t)}_\bx(U_0)$, resulting from integration of the fermions along the timelike chain at fixed spatial coordinates $\bx$ in (\ref{dmuxact1}), can also be 
explicitly evaluated  
in the gauge (\ref{Polgauge2}). One readily obtains \cite{T1}: 
\bea 
{\rm Det}{\bf M}^{(t)}_\bx(U_0) & = & {\rm Det} C^{-1}(\Theta(\bx)) 
  \nonumber \\
 & = &  2^{-\nu L} e^{ \nu\mu L}  \left( \prod_{a=1}^{N_c} \, \Big[ 1 + e^{\D -i\theta_a}
 e^{\D - \mu L}\Big]^2 \right)^{N_f} \, . \label{Det} 
\eea
Note that in the case of $N_c=2$ one has $\theta_1=-\theta_2\equiv \theta$ and (\ref{Det}) becomes 
\beq 
{\rm Det} C^{-1}(\Theta(\bx)) =   2^{-\nu (L-1)} [ \cos\theta + \cosh L \mu]^{2N_f}
\, , \label{Detsu2} 
\eeq
which is indeed real and non-negative.

\section{The cluster expansion  \label{S2}}
\setcounter{equation}{0}
\setcounter{Roman}{0}

The expansion is now generated by expanding in (\ref{meas2}) the products over the plaquettes and the  spacelike bonds in the measure of the timelike $d\mu_{\bf x}$'s  
given by (\ref{dmux}), (\ref{z}) and (\ref{dmuxact2}). In other words, we will perform: (i) 
a strong coupling expansion of the gauge field plaquette interaction together with (ii) a fermionic hopping expansion in the spacelike directions with the spacelike hops connected in the timelike directions by the propagators (\ref{prop3a})-(\ref{prop3b}).

\subsection{Diagrammatics of the expansion} 
Expanding the products 
gives
\beq 
\tilde{Z}_\Lambda = \sum_B \int \prod_{\bx \in \Lambda_s} d\mu_{\bx} 
\prod_{b_s \in B} dU_{b_s}\prod_{p\in P} f_p \prod_{b_s, l, \alpha \in B} f_{b_s}^{l,\alpha} \label{Zexp1}
\eeq
The sum is over all subsets $B$ of the set of $2\nu$ copies of the set of spacelike bonds, and over all subsets $P$ of plaquettes in $\Lambda$.  
Each such $B$ may be decomposed into a number of connected components where connectivity for 
spacelike bonds is defined as follows: two elements of a set $B$ are connected if
they may be joined by a sequence of time-like bonds. Two plaquettes in a set $P$ are said to be connected if they share a bond in their boundary. A space-like bond in a set $B$ and a plaquette in a set $P$ are said to be connected if they overlap in the plaquette boundary. 
Corresponding to this decomposition into connected sets, and after carrying out fermion and gauge field integrations, each term in (\ref{Zexp1}) decomposes into 
a product or a sum of products \cite{F3} of factors, each factor being the value of a connected diagram. 
It is important that an equal mod $N_c$ number of $U_{b_s}$ and $U^\dagger_{b_s}$'s must reside on each $b_s$ in order to obtain a nonvanishing result upon performing the $U_{b_s}$ integrations. This is, of course, a characteristic feature  of any strong coupling plaquette and/or fermion hopping expansion which determines the structure of the resulting diagrams.

\begin{figure}[t]
\qquad \quad \includegraphics[width=12.5cm]{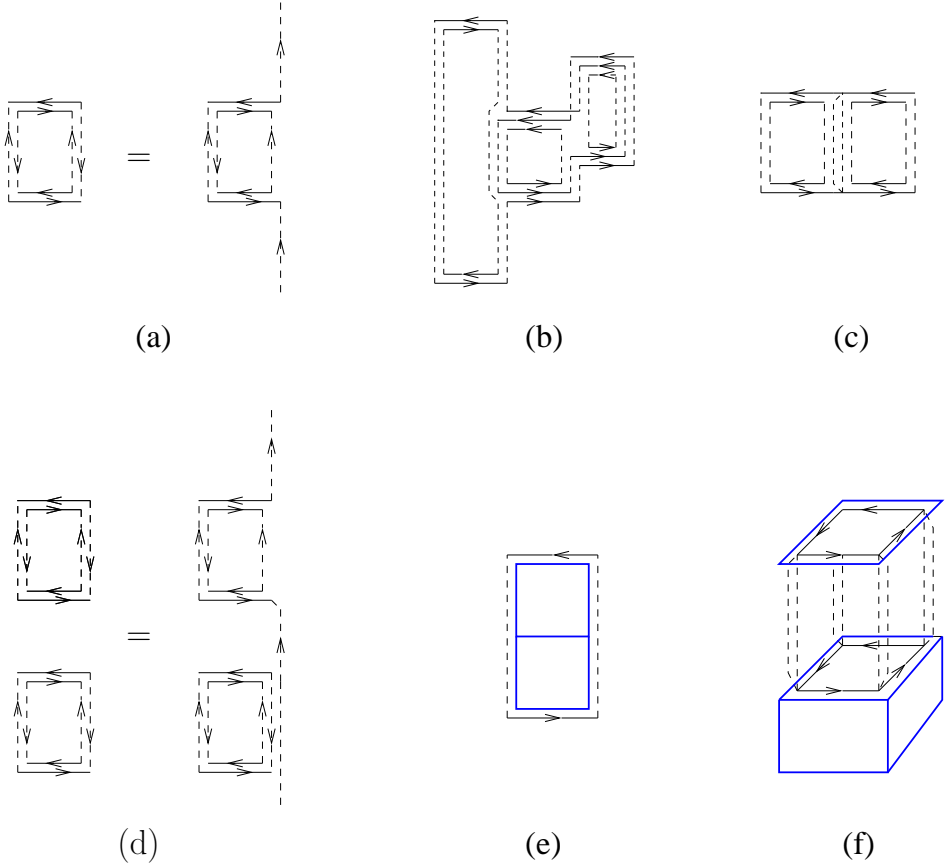} 
\caption{(a) - (d): examples of connected diagrams built of fermion hopping $f_{b_s}$ spacelike bond 
factors (solid lines) connected by timelike propagators (broken lines). 
(e) - (f): examples of general diagrams involving $f_{b_s}$ and gauge field  $f_p$ plaquette factors 
(blue). An equal mod $N_c$ number of $U$ and $U^\dagger$ must occur on every spacelike bond ($N_c=3$ in (b)). To avoid cluttering, $U$ vs $U^\dagger$ are indicated by arrows only on $f_{b_s}$ spacelike bonds, not on plaquette boundary bonds; and 
the direction (timelike forward or backward) of propagators is indicated only in (a) and (d) showing the equivalence due to timelike periodicity (cf. text). 
 \label{cpcr-hF1} }
\end{figure}

Consider first diagrams constructed solely from $f_{b_s}$ factors. 
Any such  connected diagram $\gamma$ consists of a number of  spacelike bonds connected by $1$-dimensional chains of timelike bonds corresponding to propagators, and with 
the number of $f_{b_s}$ on each spacelike bond $b_s$ such that   
equal mod $N_c$ number of $U_{b_s}$ and $U^\dagger_{b_s}$'s reside on $b_s$. 
Examples are shown in Fig. \ref{cpcr-hF1} (a) - (d). Diagrams constructed solely from plaquettes must, by the same token, form surfaces that are either closed or have boundaries consisting only of time-like bonds.\footnote{Timelike bonds may give non-vanishing contributions as free boundary (i.e. unpaired) bonds because of the ${\rm Det} C^{-1}(\Theta(\bx))$ measure factors. 
As seen from (\ref{Det}), however, such diagrams are always suppressed by extra powers of $e^{-\mu L}$ relative to those forming complete surfaces and thus make neglibible contribution to the overall sum over plaquette-generated diagrams.\label{F1}} In particular, (parts of) polymers residing wholly within a fixed time lattice slice must form closed surfaces.    Finally, a 
general diagram $\gamma$ consists of a number of spacelike bonds originating from fermionic $f_{b_s}$ factors, connected by $1$-dimensional timelike propagators, and a number of plaquettes from $f_p$ factors, the total number of $U_{b_s}$ and $U^\dagger_{b_s}$'s contributed on each bond $b_s \in \gamma$ 
from these factors being equal mod $N_c$. Examples are shown in Fig. \ref{cpcr-hF1} (e) - (f), where 
$f_{b_s}$ factors tile free plaquette boundary spacelike bonds (as, e.g., due to the missing top plaquette in the cube in (f)). 

\begin{figure}[ht]
\begin{center}
\includegraphics[width=0.35\textwidth]{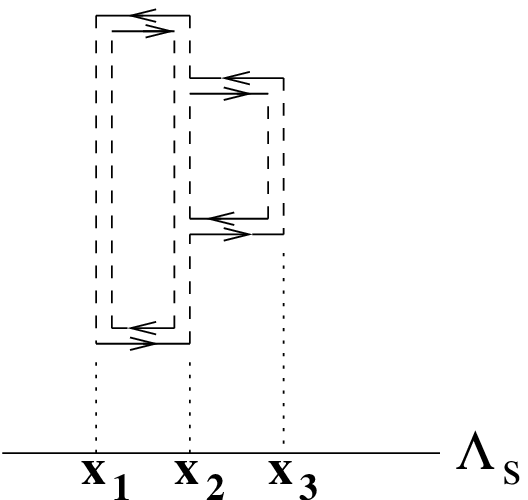}
\end{center}
\caption{The set $\{\bx \in \gamma\}$, in this example $\{\bx_1, \bx_2,\bx_3\}$, belonging to a given connected diagram $\gamma$ (cf. text).    \label{cpcrF2}}
\end{figure}

Each such connected diagram  $\gamma$ defines a `polymer'. 
The value of the diagram $\zeta(\gamma)$ is the activity of the polymer $\gamma$: 
\bea 
\zeta(\gamma) \!\!\!
& = &  \!\! \!( \prod_{{\bf x}\in \gamma} z^{-1}) \int \prod_{{\bf x}\in \gamma} dU_0(\Theta({\bx})) \prod_{{\bf x}\in \gamma}
{\rm Det} C^{-1}(\Theta(\bx)) \prod_{{\bf x}\in \gamma} \prod_{(\tau_i \tau_j)}  C_{\tau_i\tau_j} (\Theta(\bx))  I_\gamma(\{ \Theta\})  
\, .  \label{polactiv2}  
 \eea
Here $\{\bx \in \gamma\}$ denotes the set of spatial sites obtained by projecting $\gamma$ 
onto a spacelike slice $\Lambda_s$ (cf. Fig. \ref{cpcrF2}). 
In (\ref{polactiv2}) 
$(\tau_i\tau_j)$ denotes pairs of points along a timelike bond chain at $\bx$ connected by propagators. $I_\gamma$ is a $N_c, N_f$-dependent function of $\{\Theta(\bf x)\}$  resulting from the  fermion and $U_{b_s}$ integrations in the $f_{b_s}$ and $f_p$ factors used to build the diagram.

Note that diagrams such as (c), (d) in Fig. \ref{cpcr-hF1} cannot be factored in the product of two diagrams (a) because of the integration over the background gauge field $\Theta({\bf x})$ residing in propagators having common spatial site ${\bf x}$ (cf. (\ref{polactiv2})). 
As noted above  propagation backward in time from $\tau^\prime$ to $\tau$ is equivalent to propagating forward from $\tau^\prime$ winding around the periodic time direction  to $\tau$ 
(cf. (\ref{prop3a}), (\ref{prop3b})), which is indicated diagrammatically for one of the propagators in Fig. \ref{cpcr-hF1} (a) and (d). This direction equivalence may in fact be used to diagrammatically make explicit the ``connectedness"  of, for example,  Fig. \ref{cpcr-hF1} (d).

The expansion (\ref{Zexp1}) then has the form: 
\beq
\tilde{Z}_\Lambda = 1 + \sum_{k=1}^\infty  \; {1\over k!} \sum _{(\gamma_1, \ldots, \gamma_k) \in {\cal D}_k } \; \prod_{i=1}^k \zeta(\gamma_i)  \; ,\label{tildeZ3}
\eeq
where ${\cal D}_k$ denotes the set of all sets $(\gamma_1, \gamma_2, \ldots, \gamma_k)$ of $k$ disjoint polymers. In (\ref{tildeZ3}) we sum over all ordered sequences of polymers and, correspondingly, divide by $k!$. 

\begin{figure}[ht]
\begin{center}
\vspace{0.4cm}  
\includegraphics[width=0.6\textwidth]{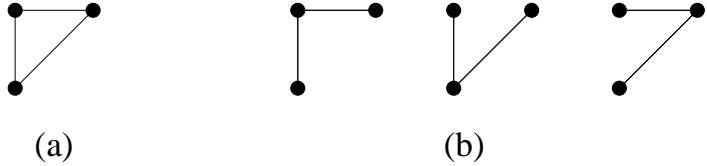}
\end{center}
\caption{(a) The graph $G(X)$ of a  cluster $X=\{\gamma_1,\gamma_2,\gamma_3\}$  consisting of three mutually intersecting polymers. (b) The set of all connected (proper) subgraphs  on $X$. \label{cpcrF3}} 
\end{figure}

The standard polymer expansion of the log of the partition function $\tilde{Z}_\Lambda$ is now obtained as follows. A set $X = (\gamma_1, \ldots, \gamma_k)$  
of $k$ polymers (not necessary distinct) will be called a $k$-polymer ($k$-cluster).  
To each such $X$ associate a $k$-vertex graph G \footnote{Here the standard graph theory definition of ``graph" is used, i.e.,  a set of vertices connected by lines (edges), with any two vertices connected by at most one line. \label{F2}} in the following manner. Each vertex in $G$ represents one $\gamma_i \in X$, and two vertices representing $\gamma_i$ and $\gamma_j$ are connected by a line if they intersect (when identified with subsets of the lattice $\Lambda$ as described above). An example is shown in Fig. \ref{cpcrF3}(a). 
 A  polymer occurring with multiplicity $m$, ($1\leq m\leq k$), in $X$ contributes $m$ vertices pairwise connected with lines (intersects itself). 
Thus the set ${\cal D}_k$ of all disjoint $k$-polymers in (\ref{tildeZ3}) 
is the set of all $X$ whose associated graphs consist of $k$ disconnected vertices (no lines). 
The set of all connected $k$-polymers, i.e. those $X$ 
whose associated graph is a path-connected $k$-vertex graph, will be denoted by $\C_k$. 
E.g., the cluster in Fig. \ref{cpcrF3}(a) belongs to $\C_3$. 
Then one has (\cite{Cam}, \cite{BGJ}) 
\beq 
\ln \tilde{Z}_\Lambda = \sum_{k=1}^\infty  \; {1\over k!} \sum _{X \in {\C}_k} \; q(X) \prod_{\gamma \in X} \zeta(\gamma)  \; , \label{logZ1}
\eeq
where the index $q(X)$ of the connected cluster $X$ is given by 
\beq
q(X) = \sum_{G_c\; {\rm on}\; X}  (-1)^{l(G_c)} \,. \label{logZ2}
\eeq
In (\ref{logZ2}) the sum is over the graph and all connected subgraphs $G_c$ on $X$, and 
$l(G_c)$ denotes the number of lines in $G_c$. Thus, e.g., the cluster in Fig. \ref{cpcrF3} has 
index $q(X) = -1 + 3\times 1= 2$.

Expectations (\ref{exp1}) are easily expressed in this formalism as follows. 
Let $Z_\Lambda[{\cal Q}]$ denotes the partition function with the insertion of any operator 
${\cal Q}$ in the measure. Given some operator ${\cal O}$ of interest, consider the 
partition function $Z_\Lambda[1 + \lambda {\cal O}]$. Its expectation is then given by 
\beq 
\vev{{\cal O}} = {Z_\Lambda[{\cal O}] \over Z_\Lambda} = 
{d\over d\lambda}\ln \tilde{Z}_\Lambda [(1+\lambda {\cal O})] 
\big\arrowvert_{\lambda=0} \,. \label{exp}
\eeq
Upon expansion of $\ln \tilde{Z}_\Lambda $ in (\ref{exp}) 
the derivative  at $\lambda=0$ picks out only terms linear in $\lambda$ resulting in a polymer expansion (\ref{logZ1}) but where each term consists of a connected $k$-cluster containing 
exactly one polymer with non-empty intersection with ${\cal O}$.

\subsection{Convergence} 
Having formulated our expansion as a polymer expansion for the logarithm of the partition function (\ref{logZ1}) or  for observable expectations (\ref{exp}) we now proceed to examine its  convergence. This may be done by an application of known convergence criteria \cite{Cam}, \cite{BGJ}, \cite{BFP} for such expansions. 
For translation invariant systems, the following convergence criterion holds: 
polymer expansions (\ref{logZ1}) and  (\ref{exp}) converge absolutely and uniformly if \cite{BGJ}
\beq
Q\equiv  \sup_x \sum_{\gamma \ni x} \,|\zeta(\gamma)| \, e^{|\gamma|} \;  < \; 1  \,.  \label{convcrit1}
\eeq
In (\ref{convcrit1}) the sum is over all polymers $\gamma$ holding fixed a site $x\in \gamma$.

It is easily seen that the value $\zeta(\gamma)$ of each connected diagram $\gamma$ in our expansion is actually proportional to $\exp (-n\mu L)$, where $n$ equals the number of `backward' moving propagators in the diagram - equivalently, as noted above, every backward moving propagator connecting two sites may be viewed as connecting the two sites by winding around the periodic time direction in the opposite, i.e. forward direction.  
So its value depends on the net quark winding number in the positive time direction as 
expected in the presence of quark (as opposed to antiquark) chemical potential. 
Keeping track of such directions, however, makes estimating the number of diagrams, and its $L$ dependences, quite  cumbersome. It is therefore convenient to proceed to bound the activities and number of diagrams in the manner of \cite{T1} at the cost of some overestimating, which, however, is not essential for the purposes of just establishing convergence of the expansion.

Now, from (\ref{prop3a}) - (\ref{prop3b}) the propagator matrix elements are bounded by 
\beq 
|C(\tau^\prime - \tau^{\prime\prime}, \Theta(\bx))| \leq {2\over [1-e^{\D - \mu L}] }\,  e^{\D -\mu \tau}  
\, . \label{propbound}
\eeq
We take $\tau$ on the r.h.s. to be  the ``periodic" (shortest) timelike distance between $\tau^\prime$ and $\tau^{\prime\prime}$ to bound each propagator in diagrams 
such as those in Fig. \ref{cpcr-hF1} (a) - (c), (e) - (f). In the case of a diagram $\gamma$  consisting of two or more separate but still ``connected" pieces, according to our notion of connectivity above, as e.g. in Fig. \ref{cpcr-hF1} (d), it is convenient to adopt the following. For each ${\bf x} \in \gamma$  common to two or more pieces, designate 
a reference fixed site on one of the pieces. Then 
take the bounding $\tau$ in (\ref{propbound}) for any `backward'  propagator at that ${\bf x}$ in the other pieces to be the periodic distance to the reference site. 
E.g. the two bottom sites in the bottom diagram in Fig. \ref{cpcr-hF1} (d) may serve as reference sites for the two  backward propagators, at the respective ${\bf x}$, in the top diagram.

 Also, from (\ref{Det}) one has 
\beq 
2^{-\nu L} e^{ \nu\mu L} \Big[ 1 -  e^{\D - \mu L}\Big]^{2\nu}
\leq   |{\rm Det} C^{-1}(\Theta(\bx))| \leq   2^{-\nu L} e^{ \nu\mu L} \Big[ 1 +  e^{\D - \mu L}\Big]^{2\nu} \,. \label{Detbound}
\eeq

Consider now a polymer $\gamma$ built out of a set of plaquettes $\gamma_p$  and a set of spacelike bonds $\gamma_b$ with bonds of multiplicity $m$ being counted $m$ times. Note that if $b\in \gamma_b$ is not in the boundary of one of the plaquettes in $\gamma_p$ then  necessarily 
its multiplicity $m\geq 2$  because of the mod $N_c$ equal number of $U_b$ and $U_b^\dagger$'s integration constraint. By the same token every plaquette in $\gamma_p$ cannot have free spacelike bonds in its boundary, i.e. every spacelike bond in its boundary must be either shared with another plaqutte in $\gamma_p$ or overlap with a bond in $\gamma_b$. 
Now, $U_{b_s}$, $U_{b_s}^\dagger$ are bounded by unity in color space. 
There are at most $\nu$ choices for connecting a $\bpsi$ or $\psi$ at a site $(\bx, \tau)$ on the boundary of one bond to  a $\psi$ or $\bpsi$ at a boundary site $(\bx, \tau^\prime)$ of another bond in the set $\gamma_b$ via a propagator $C(\tau-\tau^\prime)$. There are $|\gamma_b|$ propagators in the diagram, since each propagator connects two bonds but each bond has two boundary sites. Also, note that only sites separated by an odd number of bonds in the timelike direction can be so connected.  

From (\ref{z}), (\ref{polactiv2}) and (\ref{propbound}), (\ref{Detbound}) then one has 
\beq 
|\zeta(\gamma)| \leq (\nu)^{2|\gamma_b|} ||f_p||^{|\gamma_p|} \left[{1 +  e^{\D - \mu L} \over 1 -  e^{\D - \mu L}}\right]^{2\nu |\bx|}   \left[{2 \over 1 -  e^{\D - \mu L}}\right]^{ |\gamma_b|} 
\prod_{{\bf x}\in \gamma} \prod_{(\tau_i \tau_j)}   e^{-\D \mu \tau_{i,j}} \,  \label{activbound}
\eeq
with each $\tau_{i,j}$ for the propagator joining $\tau_i$ and $\tau_j$ chosen as explained following 
(\ref{propbound}) above. 
To next sum over all such polymers, made of $|\gamma_b|$ spacelike bonds and $|\gamma_p|$ plaquettes and attached to a fixed site, we  sum over all possible timelike separations of bonds in $\gamma_b$
and over all possible configurations of the bonds and plaquettes with the notion of connectivity defined above. Now 
\beq
\sum_{\tau=1}^{L/2} e^{\D - \mu|\tau|} <  e^{\D - \mu} [1 + \sum_{\tau=1}^{L-1} 
e^{\D - \mu|\tau|}]=  e^{\D - \mu}\left[{ 1- e^{\D - \mu L} \over 1- e^{\D - \mu}}\right] \,,
\label{propsum} \nonumber 
\eeq
whereas it is easily seen that the number of possible spacelike bond configurations is bounded by 
$(2d)^{2|\gamma_b|}$ and that of the possible plaquette configurations by $(2d-2)^{4|\gamma_p|}$. 
Combining with (\ref{activbound}) and noting that $|\bx|\leq |\gamma_b|/2$ one finally has 
\beq 
Q <\sum_{|\gamma_p|=0}^\infty  \sum_{|\gamma_b|=1}^\infty ||f_p||^{|\gamma_p|} (2d-2)^{4|\gamma_p|} K^{|\gamma_b|} (e^\mu -1)^{-|\gamma_b|} e^{|\gamma_b|+|\gamma_p|} = 
{eK\over e^\mu -1 - eK}\, {1\over 1-\epsilon}\, ,  \label{Qsum} 
\eeq 
where 
\beq 
K\equiv 2\nu^2 (2d)^2 \left[ { 1+ e^{\D - \mu L} \over 1 - e^{\D-\mu L} } \right]^\nu 
 \label{Kdef} 
\eeq
and 
\beq 
\epsilon \equiv e||f_p||(2d-2)^4 \, . \label{epsilondef}
\eeq
In (\ref{Qsum}) the sum is over all polymers attached to a fixed site and built out of at least 
one bond and any number of plaquette factors. These are the polymers 
contributing to any fermion operator expectations (\ref{exp}). 

There are also the polymers containing only plaquettes attached to 
the fixed site $x$, i.e. all polymers 
resulting solely from the expansion of the gauge field action - a minimal such polymer consists of a 3-cube. Their sum $Q^\prime$ is  bounded by 
\beq 
Q^\prime  <\sum_{|\gamma_p|=6}^\infty   ||f_p||^{|\gamma_p|} (2d-2)^{4|\gamma_p|} 
 e^{|\gamma_p|} = 
{\epsilon^6 \over 1-\epsilon}\, .  \label{Qprimesum} 
\eeq

We require  then that 
\beq 
Q <   {eK\over e^\mu -1 - eK}\, {1\over 1-\epsilon}  < 1 \, ,  \label{C1} 
\eeq 
and   
\beq
Q^\prime < {\epsilon^6  \over 1-\epsilon}  < 1   \;. \label{C2}
\eeq
(\ref{C1}) and (\ref{C2}) give the convergence conditions for our expansion. 
We thus conclude that: {\it for any spatial dimension $d\geq 1$ the expansions (\ref{logZ1}) and (\ref{exp}) converge absolutely and uniformly in $|\Lambda_s|$ 
if }
\beq 
2e\nu^2 (2d)^2 \left({2-\epsilon \over 1-\epsilon}\right) < [\tanh(\mu /2a_{\tau}T)]^\nu\,( e^{\D \mu} - 1)
\,. \label{convcrit2}
\eeq
{\it for any $epsilon$ satisfying (\ref{C2}), e.g.} $\epsilon \leq 3/4$. The condition on $\epsilon$ gives the convergence condition on $\beta$. 
In particular, a sufficient condition on $\mu$ for convergence at  any temperature ($2 \leq L < \infty$) is  given by 
\beq
\ln\left[1 + 2e\nu^2 (2d)^2  \left({2-\epsilon \over 1-\epsilon}\right)    \right] <  
\mu + \nu \ln  \tanh \mu \,. \label{convcrit3}
\eeq 
with, say, $\epsilon \leq 3/4$.

\begin{figure}[ht]
\begin{center}
\includegraphics[width=0.4\textwidth]{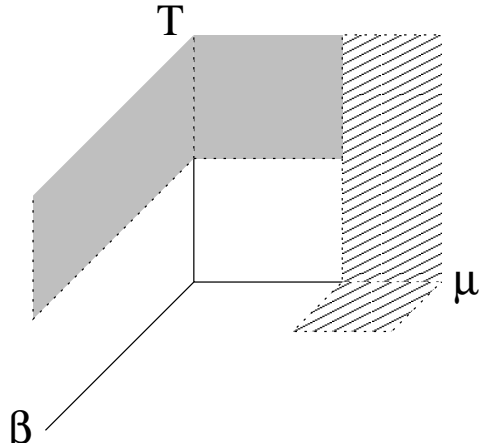}
\end{center}
\caption{ Striped region of convergence (chiral symmetry)  given by (\ref{convcrit3}) in text. 
The complete chirally symmetric region is expected to include the shaded area in the $\mu - T$ plane connecting to the region of low $\mu$ and high $T$. Chiral symmetry at $\mu=0$ and high $T$ for all $\beta$ was established in \cite{TY}. \label{cpcr-hF4}} 
\end{figure}

As already noted the value of an individual diagram is substantially overestimated  
by bounding all propagators in it by the `shortest distance' bound (\ref{propbound}) 
without proper accounting of their time directions. 
We have also summed independently over propagator distances without regard to constraints due to the fact that they connect sites that are boundary sites of spacelike bonds, i.e constrained in pairs.   
Furthermore, in our estimates we have, for the most part, simply 
ignored the spacelike bond multiplicity constraints, as well as the restriction to odd timelike separations. Also, we have ignored all factors of inverse powers of $N_c$ resulting from the spacelike gauge field integrations. 
The substantial bound improvements from taking all these into account, however, do not change the qualitative picture that follows from our result and depicted in Fig. \ref{cpcr-hF4}. This picture is further discussed in section 4.

\subsection{Chiral symmetry} 

Within its radius of convergence a cluster expansion gives complete characterization of a phase. It exhibits, in particular, all symmetries present in the infinite volume limit. 
In the present case, an immediate consequence of the convergence of our cluster expansion (\ref{logZ1}), (\ref{exp}) is the existence of chiral symmetry within the region of its convergence. 
Indeed, the expectation of any local chirally non-invariant fermion operator ${\cal O}(x)$, such as, e.g., $\bpsi(x)\psi(x)$, vanishes identically  term by term in the expansion (\ref{exp}) by the chiral invariance of the measure $d\mu_\bx$. Correlation functions $\vev{{\cal O}(x) {\cal O}(y)}$ then receive vanishing contributions from any polymers consisting of two disjoint clusters connected to sites $x$ and $y$, respectively. Nonvanishing contributions  arise only from polymers intersecting  both sites $x$ and $y$. A straightforward consequence of this fact \cite{T1} is that, 
within the expansion convergence radius, 2-point correlations are absolutely and uniformly bounded at any temperature from above by 
\beq 
\left|\vev{{\cal O}(x) {\cal O}(y)}\right|  < {\rm C_0} 
{\rm C}^{- |x-y|}  \, , \label{corrbound}
\eeq
where $C_0, C$ are space dimension-dependent constants and $|x-y|$ is the minimum number of bonds connecting the two sites. In other words, there is clustering of 2-point corellations: 
\beq
 \lim_{|x-y|\to \infty}\lim_{|\Lambda_s| \to \infty} \vev{{\cal O}(x) {\cal O}(y)}  \to 0 \,.  
 \label{corrcluster}
 \eeq
 In the same manner, all higher correlation  functions of any chirally non-invariant fermion operators exhibit exponential clustering for large separations, i.e., there is no spontaneous breaking of the global chiral symmetry.

\section{Discussion \label{S4}}
\setcounter{equation}{0}
\setcounter{Roman}{0}

We have shown the existence of a chirally symmetric phase at sufficiently large chemical potential $\mu$ and sufficiently large gauge coupling at any temperature $T$ in $SU(N_c)$ with $N_f$ flavors of massless staggered fermions. This is an exact lattice result obtained by a convergent cluster expansion in the infinite volume limit. 
Chiral symmetry in $SU(N_c)$ gauge theories at vanishing $\mu$ and $T$ is  of course spontaneously broken, a long-studied phenomenon by MC simulations and many other methods. 
In the case of strong coupling it was 
established analytically in lattice gauge theory at large $N_c$ and fixed $N_f$ by resumed hopping expansions and other techniques long ago \cite{BBEG}. Our result then implies the existence of a chiral symmetry restoring phase transition at some critical $\mu$, as indeed seen in the strong coupling numerical and mean field studies cited in section 1. 

The chirally symmetry phase at strong coupling, high $\mu$ and low $T$ is a confined phase with a parity doubled spectrum of mesons and baryons. It is continuously connected, as we established in this paper, to the high $\mu$ and high $T$ region. This latter region is expected in turn to be continuously connected to the low $\mu$, high $T$ deconfined region through the shaded area in  Fig. \ref{cpcr-hF4}. It would appear quite feasible to show this by further application of the methods used here. To do this one would first need to 
allow independent lattice spacings $a_s$ and $a_\tau$ in the spacelike and timelike directions. 
In this paper the ratio $(a_\tau/a_s)$ was set equal to unity at the outset. Allowing general values for it, however, provides essential convergence factors at low $\mu$ and high $T$ (cf. \cite{TY}). Secondly, one would need to bound the gauge field $\Theta$ dependence at small $\mu$ more precisely than in (\ref{propbound}) and (\ref{Detbound}). These steps should allow one to connect to the convergence region obtained in \cite{TY}. 
The remaining unshaded region in the $\mu$ - $T$ plane in Fig. \ref{cpcr-hF4} contains the usual hadronic  phase. The resulting picture then resembles that of ``quarkyonic'' matter 
as discussed in \cite{McLP}. 
In this paper, working at strong coupling, we have extended this picture to a finite region in $\beta$. 

The question now is what, if any, qualitative changes are encountered in the large $\beta$ (weak bare coupling) regime. 
At vanishing $\mu$ and high $T$ the existence of a chirally symmetric phase was rigorously established for all $\beta$ in \cite{TY},  as indicated in Fig. \ref{cpcr-hF4}.  
The region of (asymptotically) large $\mu$ and large $\beta$ at low $T$ is the regime where 
a color superconductivity phase via the formation of color-flavor locking (CFL) condensates has been extensively discussed \cite{ASRS}. Such condensates generally also entail chiral symmetry breaking. 
This implies that, if such a phase is realized, there must be a phase transition in the lattice gauge theory as a function of the bare gauge coupling at fixed large $\mu$ and low $T$.

The argument for a CFL phase originates at large $\mu$, at which scale the running coupling is weak 
and the usual Cooper pairing argument near the Fermi surface is invoked. In the case of three massless or light fermion flavors with $N_c=3$ the CFL phase is argued by ``hadron-quark 
continuity" \cite{ASRS} to extend to and in fact continuously connect to the hypernuclear matter region of the phase diagram at $\mu$ values corresponding to intermediate or strong  coupling. In the case of two massless flavors (sufficiently heavy strange quark), or other $N_f$ and $N_c$ choices,  other phases, such as the chirally symmetric 2CS phase, may intervene at intermediate $\mu$. 
In this connection recall that $N_f$ flavors of massless staggered fermions correspond to $4N_f$ flavors  in the continuum. The argument for CFL phases with $N_f=2,3$, however, can be equally  
well applied to cases with $N_f \geq 4$ \cite{ASRS}. If $N_c=3$ and $N_f$ is a multiple of three, for example, the order parameter takes the form of multiple copies of the $N_f=3$ order parameter, i.e. blocks of three flavor-color locking. 

All this means that there is no immediate apparent contradiction between the exact 
large $\mu$, small $\beta$ lattice results presented here and the existence of CFL phases. 
It is, however, noteworthy that at high $\mu$  
no chiral symmetry breaking occurs at strong bare coupling for any choice of $N_c$, $N_f$.  
The only firm consequence that may be drawn at this juncture is that, as already pointed out, if the predictions of a CFL phase for some choices of $N_c, N_f$ are indeed valid, the corresponding lattice theory must exhibit a phase transition as $\beta$ is increased. 
It would clearly be very interesting if the exact lattice results could be extended into the large $\mu$, large $\beta$ region.

This might be possible in certain cases. 
As it is well known the fermion determinant is generally complex for $N_c \geq 3$. With the partial exception of two-color QCD,  most investigations, understandably, concern the physical case of $N_c=3$ rather than other choices of number of colors. 
The real or complex nature of the fermion determinant is not of relevance 
within the cluster expansion approach used here since it is based on expansion of the measure in a 
convergent series - this is in fact one of its great virtues. 
In their present form, however, our expansions cannot be used at weak coupling. It may be possible though to rearrange these expansions so that they can be used at large $\beta$ provided that any unexpanded (resumed) parts of the measure are real positive. This seems more feasible in the case of even number of colors and we hope to pursue it elsewhere.

The number of fermion flavors $N_f$ is another relevant parameter for the presence or absence of 
chiral symmetry. 
At $T=0$ and $\mu=0$ chiral symmetry was recently found to be restored even in the infinite coupling limit provided the number of flavors becomes large enough (at fixed $N_c$). 
This surprising result was obtained by numerical simulations for $N_c =3$ \cite{dFKU}, and, more generally, for $SU(N_c)$ by resummations of fermion hopping expansions \cite{T2}.  
The $N_f$ dependence in the emergence of chiral symmetry at finite $\mu$ has yet to be investigated. Note that 
in our estimates for $\mu$ such that chiral symmetry is present (cf. (\ref{convcrit2})) the flavor dependence is contained in $\nu$. These estimates were obtained by bounding the absolute value of each term in the expansion from above in order to prove its absolute convergence. 
This completely obscures any potential cancellations between classes of diagrams  as $N_f$ is varied (at fixed $N_c$). Such cancellations are precisely the mechanism found to be operative at $\mu=0$ in the restoration of chiral symmetry at large $N_f$ and strong coupling: 
subdominant graphs at small $N_f$ become dominant at large $N_f$ with signs such as to 
destroy the condensate formed by those graphs dominant at small $N_f$ \cite{T2}. 
Whether analogous cancellations occur in the presence of finite chemical potential $\mu$ remains to be seen.

Another interesting class of systems for which exact lattice results may be possible is 
obtained by adding Higgs fields. This is motivated by the following consideration. The  usual CFL phase is a Higgs phase where the gluons acquire masses through the ``spontaneous breaking" of the color gauge symmetry by the formation of the condensate. The short distance gluon interaction in an attractive channel is thereby cut-off in the infrared, thus allowing weak coupling 
computation around the putative condensed ground state. An alternative way to introduce masses for  all or some of the gluons is, of course, by coupling Higgs scalar fields, in particular one fundamental representation Higgs field, to the gauge field. 
No Yukawa couplings, which explicitly break chiral symmetry, are allowed, so that 
all interactions are chirally invariant. Admittedly, the system is no longer asymptotically free for 
low $N_f$, but 
the logarithmic rise of the gauge coupling in the UV is so slow that the system can be made weakly coupled at all scales of interest. In the presence of a quark chemical potential one may again have weak attractive interactions near the Fermi surface with suppression of infrared gluonic degrees of freedom. Actually, on the lattice, by appropriately varying the Higgs field modulus, one may probe the entire range 
from strong to weak coupling. Recall that at $\mu=0$ the `confining' strong coupling regime 
and the weak coupling `Higg phase' regime are in fact continuously connected \cite{FS} - this is a well-known exact lattice result also obtained by convergent cluster expansions. 
This type of argument can be incorporated within our type of cluster expansions at $\mu\not= 0$.  It would then certainly be worth while to investigate the possibility of further exact results for the phase diagram of such systems as a function of $\mu$, $\beta$ and the Higgs modulus.

\end{document}